\documentclass[%
 reprint,
superscriptaddress,
preprint,
 amsmath,amssymb,
 aps,
pra,
onecolumn]{revtex4-2}

\usepackage{graphicx}% Include figure files
\usepackage{dcolumn}% Align table columns on decimal point
\usepackage{bm}% bold math
\usepackage{xcolor}
\usepackage{pdfpages} % include pdfs
\usepackage{pgffor} % for loops

\makeatletter
\AtBeginDocument{\let\LS@rot\@undefined}
\makeatother

\def\supplementfilename{supplement.pdf}

\pdfximage{\supplementfilename}
\def\numbersupplementpages{\the\pdflastximagepages}

\newif\ifarXiv
\arXivtrue 
\begin{document}

%\preprint{APS/123-QED}

\title{Investigation of magnetic order influenced phonon and electron dynamics in MnBi$_{2}$Te$_{4}$ and Sb doped MnBi$_{2}$Te$_{4}$ through terahertz time-domain spectroscopy}% Force line breaks with \\

\author{Soumya Mukherjee}
\affiliation{Department of Physical Sciences, Indian Institute of Science Education and Research Kolkata, Nadia, 741246, West Bengal, India}

\author{Anjan Kumar NM}
\affiliation{Department of Physical Sciences, Indian Institute of Science Education and Research Kolkata, Nadia, 741246, West Bengal, India}

\author{Subhadip Manna}
\affiliation{Department of Physical Sciences, Indian Institute of Science Education and Research Kolkata, Nadia, 741246, West Bengal, India}

\author{Sambhu G Nath}
\affiliation{Department of Physical Sciences, Indian Institute of Science Education and Research Kolkata, Nadia, 741246, West Bengal, India}

\author{Radha Krishna Gopal}
\email{rk.gopal@mail.jiit.ac.in}
\affiliation{Department of Physics and  Material Sciences and  Engineering, Jaypee Institute of Information Technology, Sector 62, Noida, India}

\author{Chiranjib Mitra}
\affiliation{Department of Physical Sciences, Indian Institute of Science Education and Research Kolkata, Nadia, 741246, West Bengal, India}

\author{N. Kamaraju}
\email{nkamaraju@iiserkol.ac.in}
\affiliation{Department of Physical Sciences, Indian Institute of Science Education and Research Kolkata, Nadia, 741246, West Bengal, India}

\date{\today}% It is always \today, today,

\begin{abstract}
MnBi$_{2}$Te$_{4}$, the first topological insulator with inherent magnetic ordering, has attracted significant attention recently for providing a platform to realize several exotic quantum phenomena at relatively higher temperatures. In this work, we have carried out an exhaustive investigation of MnBi$_{2}$Te$_{4}$ and Sb doped MnBi$_{2}$Te$_{4}$ thin films using THz time-domain spectroscopy. The extracted real THz conductivity displays a strong IR active E$_u$ phonon absorption peak (at $\sim$1.5 THz) merged on top of the Drude-like contributions from bulk and surface electrons. The extracted parameters from the THz conductivity data fitted to the Drude-Fano-Lorentz model, show significant changes in their temperature dependence around the magnetic ordering Néel temperature of $\sim$ 25K, which is suggestive of the coupling between magnetic ordering and electronic band structure. The frequency of the E$_u$ phonon displays an anomalous blue-shift with increasing temperatures by $\sim$ 0.1 THz ($\sim$7\%) for MnBi$_{2}$Te$_{4}$ and $\sim$0.2 THz ($\sim$13\%) for Sb doped MnBi$_{2}$Te$_{4}$ between 7K and 250K. The line-shape of the E$_u$ phonon mode in Sb doped MnBi$_{2}$Te$_{4}$ shows significant Fano asymmetry compared to that of MnBi$_{2}$Te$_{4}$, indicating that Sb doping plays an important role in the Fano interference between the phonons and the electrons, in this system. These results indicate that the anomalous phonon behaviour seen in MBT arise mainly from positive cubic anharmonicity induced self energy parameter, whereas both anharmonicity and the electron phonon coupling are at play in making the relatively higher anomalous blue shift of phonons in MBST. Our studies provide the first comprehensive understanding of the phonon and electron dynamics of MnBi$_{2}$Te$_{4}$ and Sb doped MnBi$_{2}$Te$_{4}$ in the THz range using time-domain THz spectroscopy. 

\end{abstract}

\maketitle

%\tableofcontents

\section{\label{sec:level1}Introduction}

The introduction of magnetic order in topological insulators (TIs) unlocks the possibilities of investigating several novel quantum phenomena, such as quantum anomalous hall effect (QAHE) \cite{Chang2023,chang2013,deng2020}, axion insulator state \cite{Xiao2018,Chang2023} and magnetic Weyl semimetallic phase \cite{kopf2022,lee2021}. Although doping topological insulators (TIs) with magnetic impurities is often considered the most convenient method, the lack of precise control during the fabrication process can lead to spatial inhomogeneity and magnetic disorders \cite{Lee2015,Chang2023}. This can result in complex magnetic ordering, a small exchange gap, and a very low operating temperature for the realization of these quantum phenomena \cite{Hesjedal2019,deng2020,Otrokov2019prediction}.  Whereas the other means of exploring the coupling between topology and magnetism through the proximity induced effect in the heterostructure of magnetic insulators and TIs are challenging in material choice and interface fabrication \cite{Hao2019}, and are generally very weak in nature \cite{Chang2023}. These issues necessitate the development of stoichiometric magnetic TIs capable of hosting inherent magnetic ordering \cite{Otrokov2019prediction}. The recent introduction of the magnetic TI (MTI) family MnBi$_{2n}$Te$_{3n+1}$ thus marks the beginning of a new era of topological condensed matter physics. MnBi$_{2}$Te$_{4}$ (n=1) is the most important and most explored member of this MTI family \cite{Choe2021}. MnBi$_{2}$Te$_{4}$ hosts intrinsic magnetic ordering via intercalating a Mn-Te bilayer chain in a quintuple (QL) of Bi$_2$Te$_3$, which is a topological insulator \cite{deng2020,kopf2020}. It enters into the magnetically ordered phase below the Néel temperature, T$\mathrm{_N}$ $\sim$25 K \cite{deng2020,cui2019,Xu2021}, where the magnetic moments generated from Mn$^{2+}$ ions get aligned parallelly within a MnBi$_{2}$Te$_{4}$ septuple layer (SL) and couple antiparallelly with the adjacent SLs resulting in an intralayer ferromagnetic and overall A-type antiferromagnetic (AFM) order, respectively \cite{Li2019,kopf2020}.

The inherent topological characteristic of MnBi$_{2}$Te$_{4}$ enables the coexistence of gapless topological surface states (TSS) and a gapped bulk electronic band structure \cite{Hao2019,chen2019}. In the AFM phase, the intrinsic magnetic ordering disrupts the time-reversal symmetry, which is anticipated to trigger the opening of an exchange gap at the Dirac node \cite{tomarchio2022electrodynamics,he2020mnbi2te4}. However, this opening of the exchange gap
is highly sample dependent \cite{garnica2022native}, and has elicited divergent views based on their ARPES measurements\cite{Hao2019,chen2019}. Observations of both gapless \cite{chen2019,Hao2019}, and gapped \cite{Hao2019,Otrokov2019prediction} TSS have been reported following the onset of magnetic ordering. This discrepancy underscores the need for further exploration. The magnetically ordered phase also causes a band splitting of the conduction band of MnBi$_{2}$Te$_{4}$ by $\sim$50 meV ($\sim$12 THz) \cite{kopf2020,Xu2021,chen2019} as seen in the surface sensitive ARPES measurements. This splitting has been attributed to the uncompensated surface ferromagnetism \cite{chen2019}. However, bulk sensitive IR spectroscopy discards this attribution \cite{kopf2020}. Therefore, further spectroscopic investigations are essential to reveal the origin of the interaction between magnetic ordering and electronic band structure. Since the reconstruction of both surface and bulk band structure strongly influences the low energy IR spectrum \cite{kopf2020}, Terahertz time domain spectroscopy (THz-TDS) can play a vital role here. 
Moreover, THz-TDS can capture the possible occurrence of Fano interference\cite{Fano1961} between discrete bulk phonons and gapless TSS excitation revealing intriguing physics of topological materials \cite{Sim2015,park2015terahertz}. To the best of our knowledge, no THz-TDS measurements on MnBi$_{2}$Te$_{4}$ have been reported as of yet.

Fermi level of bulk MnBi$_{2}$Te$_{4}$ which is mostly found to get located in the bulk conduction band, can be tuned via substituting Bi with the Sb atoms \cite{kopf2022,Watanabe20222}.
Such Fermi level tuning can suppress the bulk carriers and provide an even more ideal platform for exploring high temperature QAHE \cite{chen2019intrinsic}. Also, MnBi$_{2(1-x)}$Sb$_x$Te$_{4}$ offers a greater possibility of realizing the Weyl semimetallic state, which remains otherwise elusive in MnBi$_{2}$Te$_{4}$ \cite{lee2021,kopf2022}. Moreover, lighter atom (Sb) substitution can also modulate the spatial overlap between bulk phonons and the TSS, which enhances the possibility of MnBi$_{2(1-x)}$Sb$_x$Te$_{4}$ being a tunable Fano interference system in the THz regime similar to the previously reported In doped Bi$_2$Se$_3$ \cite{Sim2015}.

Here, we have carried out temperature dependent THz-TDS measurements on epitaxial thin films of MnBi$_{2}$Te$_{4}$ (MBT) and Sb-doped MnBi$_{2}$Te$_{4}$ (MBST) grown via pulsed laser
deposition (PLD) technique. The extracted THz conductivity at various temperatures (7K to 250K) show a strong phonon absorption peak at $\sim$1.5 THz, which is identified as the IR active E$_u$ phonon mode \cite{kobialka2022}. Interestingly, the E$_u$ phonon mode of both MBT and MBST exhibits an unusual shift towards higher frequencies with increasing temperature. The THz conductivity have been fitted with Drude-Lorentz-Fano oscillator model, through which important parameters like plasma frequency, Drude scattering time and Fano asymmetry are extracted and investigated. 

\section{\label{sec:level2}Methods}

\begin{figure*}
\centering

     		\includegraphics[scale=0.52]{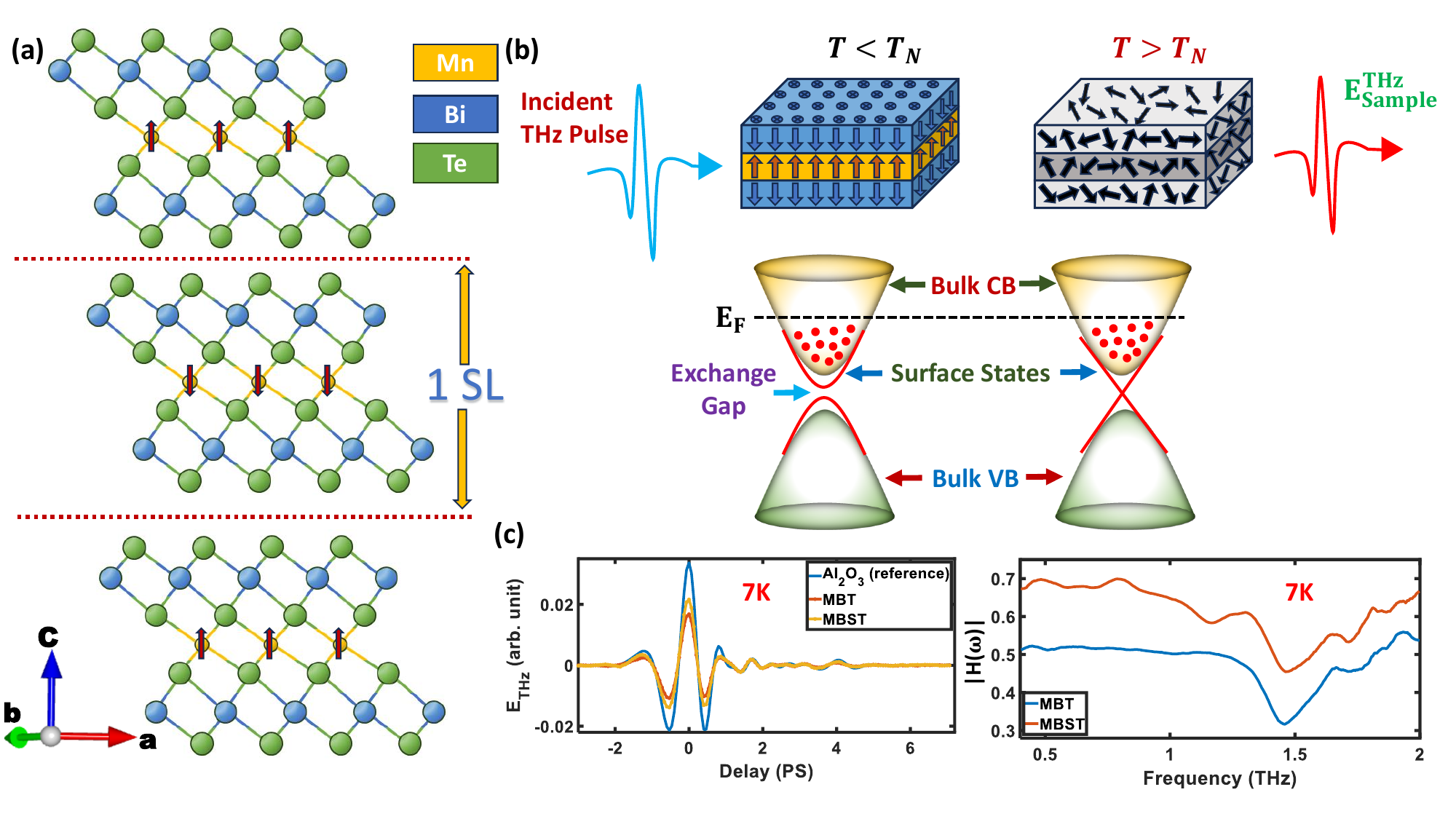}
  
   \caption{ (a) Crystal structure of MBT with AFM coupling between adjacent Mn layers. (b) Schematic of a 3 SL sample showing alignment of the magnetic moments of Mn atoms below and above magnetic ordering temperature (T$\mathrm{_N}$), and corresponding electronic band structure of MBT, showing gapped and gapless surfcae state, respectively. (c)
   Recorded THz time domain data of sapphire (Al$_2$O$_3$) substrate, MBT and MBST at 7K, and  THz transmission ($\lvert \mathrm{H}(\omega) \rvert$) through  MBT and MBST at 7K as a function of frequency ($\omega$/2$\pi$).}
   \label{fig:1}
\end{figure*}

\begin{figure}
\centering

     		\includegraphics[scale=0.63]{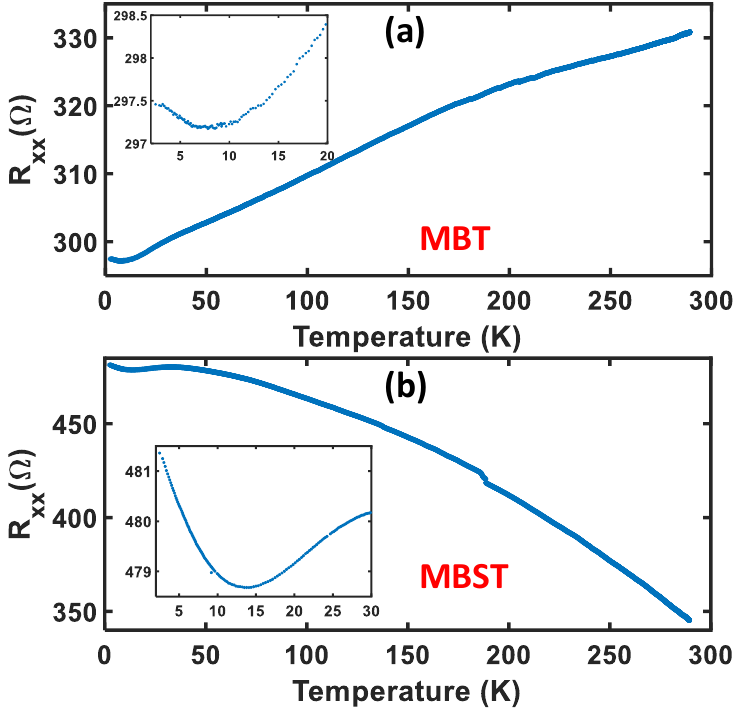}
  
   \caption{ Temperature dependence of longitudinal resistance (R$_{xx}$) of (a) MBT  and (b) MBST  thin films}
   \label{fig:2}
\end{figure}

MBT crystallizes in rhombohedral structure with R-3m space group symmetry. As shown in Fig. \ref{fig:1}(a), single SLs of MBT forms via stacking of the atoms in Te-Bi-Te-Mn-Te-Bi-Te order. The arrows placed in each Mn atom represents the direction of its magnetic moment and indicates their AFM coupling. In this work, thin films of MBT and MBST have been deposited on sapphire (Al$_2$O$_3$) substrates using PLD technique. Each film was deposited on a substrate of dimensions 10 mm$\times$10 mm$\times$430 $\mu$m. The details of the sample growing procedure using PLD technique has been discussed in the Supplemental Material \cite{supplemental}, section S1. For preparing MBST, the amount of Sb in the target pellet of PLD has been kept such that it takes the stoichiometric formula of MnBi$_{1.7}$Sb$_{0.3}$Te$_{4}$. This stoichiometric ratio of Sb is expected not to cause any considerable change in T$\mathrm{_N}$, according to what has been observed in the previously reported bulk MBST \cite{chen2019intrinsic}.

The thicknesses of the films are measured using cross-sectional SEM technique and are found to be $\sim$152 nm and $\sim$142 nm for MBT and MBST, respectively (see Supplemental Material \cite{supplemental}, Fig. S1) The characterization (XRD, SEM, EDX and Raman Spectroscopy) of the films have been discussed in the Supplemental Material \cite{supplemental}, section S2.  

The THz-TDS measurements have been carried out using a home-built transmission geometry THz-TDS setup (see Fig. S4 for schematic diagram). Our THz-TDS setup uses pulsed laser output of an amplifier (RegA 9050, Coherent Inc. (USA)) operated at 100 kHz with pulse-width of $\sim$60 fs and central photon energy of $\sim$1.57 eV, to generate and detect THz. The generation and detection of the THz pulse have been done via optical rectification and electro-optic sampling, respectively using $<$110$>$ oriented ZnTe crystals. The sample temperature was varied using a closed cycle Helium cryostat (OptistatDryBLV, Oxford Instruments). Further details of the THz-TDS setup and measurements are provided in Supplemental Material \cite{supplemental}, section S3-A.

We have recorded THz waveform of the samples along with the bare substrate in a temperature range of 7K to 250K. By taking the ratio of the Fourier transformed THz waveform of the sample ($E_{sample}(\omega)$) and the substrate ($E_{substrate}(\omega)$), we have calculated THz transmission, $H(\omega)$ (=$E_{sample}(\omega)$/$E_{substrate}(\omega)$) as shown in Fig. \ref{fig:1}(c) for 7 K temperature (THz transmission for all the temperatures are shown in Supplemental Material \cite{supplemental}, S3-B).

The AC conductivity in the THz range have been directly calculated from THz transmission ($H(\omega)$) using the thin film formula \cite{Tinkham1956}, $\sigma(\omega)=\frac{n_{substrate}+1}{Z_{0}d}\Big(\frac{1}{H(\omega)}-1\Big)$. Where, n$_{substrate}$ is the refractive index of the substrate, Z$_0$ is the free space impedance and d is the thickness of the thin film sample.

The temperature dependent electrical measurements were done via standard technique in the van der Pauw geometry. Keithley sourcemeter and voltmeter were used for the measurements. The temperature was varied via keeping the sample inside the same cryostat that has been used for THz studies.  

\section{\label{sec:level3}Results and Discussion}

The temperature dependent longitudinal resistance (R$_{xx}$) of the samples are shown in Fig. \ref{fig:2}(a) and \ref{fig:2}(b). For the temperatures above 10 K, R$_{xx}$ increases with increasing temperature for MBT (Fig. \ref{fig:2}(a)) indicating its metallic nature as observed in bulk \cite{CHANGDAR2023414799,cui2019} and flakes ($\sim$100-200 nm thick) \cite{cui2019} of single crystalline MBT . At temperature $\lesssim$10 K, R$_{xx}$ shows upturn, which suggests its nonmetallic behaviour (see inset Fig. \ref{fig:2}(a)) in very low-temperature regime. This behaviour could be attributed to the bulk carriers freezing and disorder induced electron-electron interaction \cite{CHANGDAR2023414799,Liu_2011,gopal2017topological}. Plausibly, the presence of disorder has also caused disappearance of spin fluctuation driven AFM transition peak in the R-T plot, similar to the thin film grown via co-sputtering technique \cite{Lu_2023}.

For MBST, the overall increase in R$_{xx}$ (Fig. \ref{fig:2}(b)) compared to MBT indicates bulk carrier suppression due to shift of the Fermi level towards the valance band, which is expected from Sb substitution \cite{chen2019intrinsic,Watanabe20222}. The temperature dependent trend of R$_{xx}$ of MBST mostly shows nonmetallic nature. Appearance of nonmetallic nature due to the suppression of bulk carriers via Fermi level tuning in MBST was also seen in its single crystalline form \cite{chen2019intrinsic}.  Below $\sim$30 K, R$_{xx}$ takes a downturn and shows metallic nature indicating surface-state dominated conduction process \cite{Singh_2017,gopal2017topological} due to weak anti localization of the surface electrons \cite{gopal2017topological,Taskin2012}. The metallic nature persist till $\sim$17 K, then the resistance takes an upturn which can be attributed to the two-dimensional electron-electron interaction with the disorder \cite{wang2011,Liu_2011}. A detailed temperature and magnetic field dependent transport measurements are beyond the scope of this work.

\begin{figure*}
\centering

     		\includegraphics[scale=0.74]{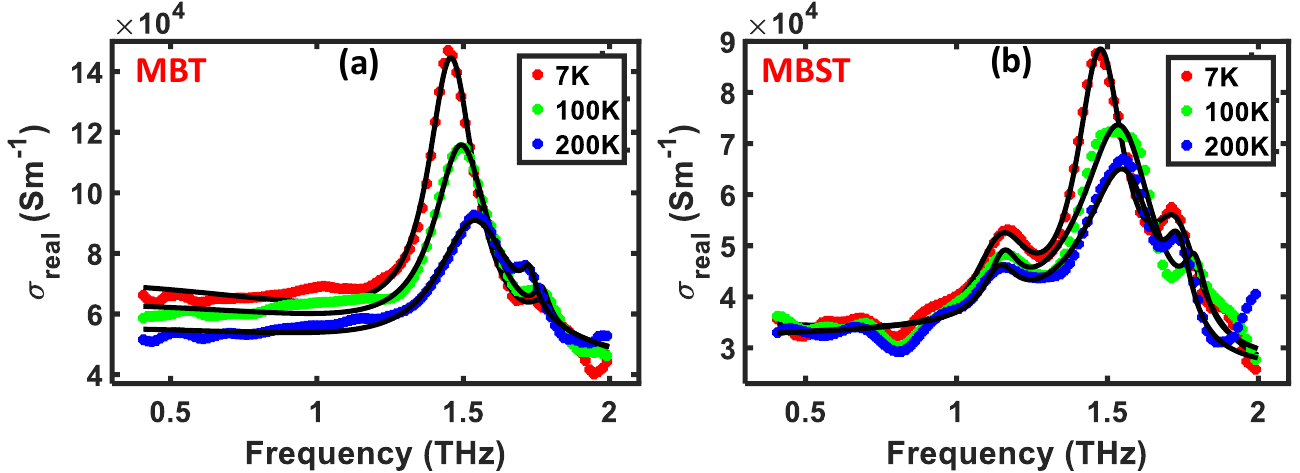}
  
   \caption{ Real part of THz conductivity at different temperatures for (a) MBT  and (b) MBST. The black continuous line represents the corresponding fit using Eq. (\ref{eq:1}).}
   \label{fig:3}
\end{figure*}

The real part of the extracted THz conductivity of MBT and MBST at different temperatures are shown in Fig. \ref{fig:3}(a) and Fig. \ref{fig:3}(b). Along with the Drude-like contributions from surface and bulk electrons, the real conductivity is found to be composed of several characteristic absorption peaks. The strongest absorption peak for both MBT and MBST is located at $\sim$1.5 THz, and can be identified as the IR active phonon mode E$_u$, which is in agreement with the theoretical estimation from first principle DFT calculations \cite{kobialka2022}. The E$_u$ phonon mode displays two significant temperature dependent features: (i) the phonon peak undergoes a blue-shift ($\sim$ 0.1 THz for MBT and $\sim$ 0.2 THz for MBST ) upon increasing the sample temperature (from 7K to 250K), which is an anomalous behaviour \cite{melnikov2022anomalous,li2020anomalous,LaForge2010} and (ii) the line-shape of the E$_u$ phonon mode appears asymmetric (mostly for MBST) indicating it to be emerging from Fano resonance \cite{Sim2015}. Fano resonance occurs due to the interference between a discrete mode and a continuum of excitations \cite{Fano1961}. In case of TIs, its appearance suggest strong coupling between discrete bulk phonon mode and transition of TSS electrons \cite{LaForge2010}.  

The measured THz conductivity have been found to fit well via one Fano and one Lorentz oscillator for MBT, and one Fano and two Lorentz oscillators for MBST (for the absorption peaks) along with the Drude contribution (for electronic response) \cite{Sim2015,wu2013sudden}. The functional form of our model is as follows,

\begin{widetext}
\begin{equation}
\sigma(\omega)=\frac{\epsilon_0 \omega_{Drude}^2 \tau_{Drude}}{(1-i\omega \tau_{Drude})}+\frac{i\epsilon_0 A_F\gamma_{Fano}\omega(i-(1/q)^{-1})^2}{\omega^2-(2\pi i f_{Fano})^2+2\pi i\gamma_{Fano}\omega}+ \sum_{k=1}^{N_L}\frac{i\epsilon_0\omega A_L^k}{\omega^2-(2\pi f_{Lorentz}^k)^2+2\pi i\gamma_{Lorentz}^k\omega}
 \label{eq:1}
\end{equation}
\end{widetext}

Here, $\tau_{Drude}$ and $\omega_{Drude}(=\sqrt{\frac{ne^2}{m^*\epsilon_0}})$ are the  Drude scattering time and plasma frequency, f$_{fano} $(f$_{Lorentz}^k$), $\gamma_{fano}$ ($\gamma_{Lorentz}^k$) and A$_F$ (A$_L^k$) are the central frequency, linewidth and oscillator strength of the Fano (Lorentz) model, respectively. Here, N$_L$ designate the number of required Lorentz oscillator, which is 1 and 2 for MBT and MBST, respectively. The parameter $\mid 1/q \mid$ is the measure of Fano asymmetry, and it represents the strength of coupling between bulk phonon and TSS. For $\mid 1/q \mid$ $\rightarrow$ 0, the asymmetry decreases and the Lorentz lineshape is restored.

\begin{figure*}
\centering

     		\includegraphics[scale=0.6]{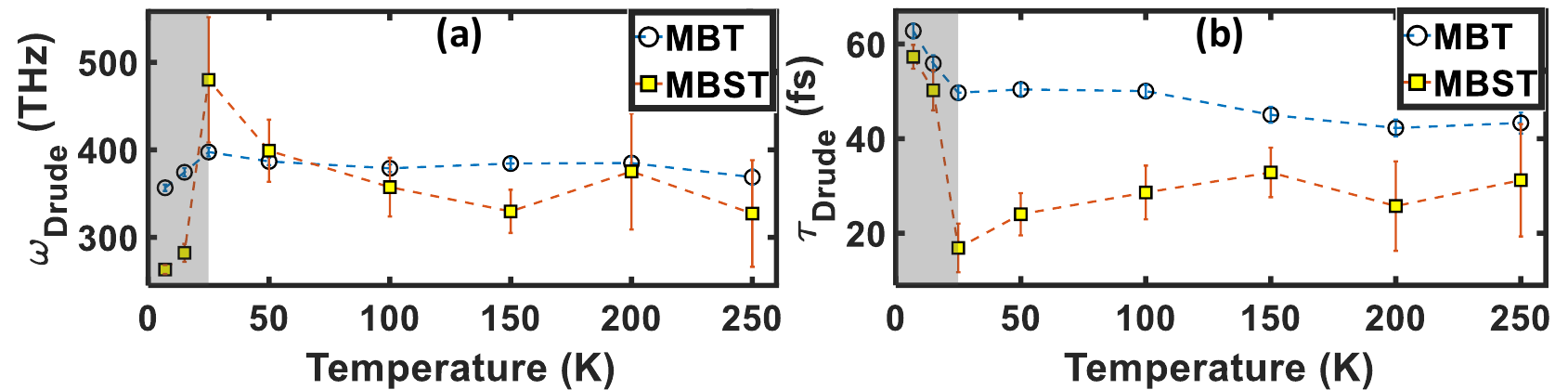}
  
   \caption{ Temperature dependence of extracted Drude parameters: (a) plasma frequency, (b) Drude scattering time  of MBT and MBST together. The grey shaded region indicates the AFM phase and the dashed lines are the guide to the eyes.}
   \label{fig:4}
\end{figure*}

The theoretical model fits the real conductivity well as shown in Fig. \ref{fig:3}(a) and Fig. \ref{fig:3}(b) (shown for few temperatures for clarity), and the corresponding R$^2$ values are listed in Supplemental Material \cite{supplemental}, Table S1 for all the temperatures.
The temperature dependence of the extracted fit parameters of Drude part for MBT and MBST together are shown in Fig. \ref{fig:4}(a)-\ref{fig:4}(b). 
The temperature dependence of $\omega_{Drude}$ shows an initial increase with increasing temperature till T$\mathrm{_N}$, followed by a moderate decrease in the rest of the temperature range for both the samples. From the IR spectroscopy of bulk MBT, Kopf et al.\cite{kopf2020} have reported a similar temperature dependent trend in plasma frequency. This observation is due to the modification of electronic bandstructure in magnetically ordered phase. In the AFM phase, an exchange gap emerges, disrupting the TSS (schematically shown in \ref{fig:1}(b)). The exchange gap (of an amount $\Delta$) modifies the energy-momentum dispersion relation, which causes the massless surface electrons (given by E= $\pm \hbar v_Fk$) to accrue an effective mass (given by E=$ \pm\sqrt{(\hbar v_Fk)^2+(\Delta/2)^2}$) \cite{kopf2022,tomarchio2022electrodynamics,Taskin2012}. The addition of effective mass 
and lifting of the topological protection results in reduced mobility, which eventually alters their conductivity contribution into a nonmetallic nature \cite{tomarchio2022electrodynamics,Taskin2012}. This phenomenon has been further corroborated through the previous IR studies of 7 SL MBT film \cite{tomarchio2022electrodynamics}, where the predominance of surface electrons is significantly higher. For the temperatures in the region above T$\mathrm{_N}$, $\omega_{Drude}$ mostly remains unaltered with increasing temperature, similar to what has been observed in 7SL MBT film \cite{tomarchio2022electrodynamics}.
In the case of MBST, the change in $\omega_{Drude}$ around T$\mathrm{_N}$ looks considerably sharper compared to that of MBT (see Fig. \ref{fig:4}(a)). The shift of the Fermi level towards the bulk valence band in MBST due to Sb doping, causes suppression of bulk carrier contribution, as suggested by the temperature dependent DC resistance measurements (see Fig. \ref{fig:2}). Thus in MBST, the suppression of bulk carriers make the THz radiation to be more sensitive to the modification in the surface states due to magnetic ordering, capturing the pronounced anomaly of $\omega_{Drude}$  around T$\mathrm{_N}$.  At low temperature regions below T$\mathrm{_N}$, the decreasing trend of $\omega_{Drude}$ is also due to the disorder induced electron-electron interaction, that causes low temperature non-metallic behaviour 
as seen in DC transport measurements (Fig. \ref{fig:2}). 

The temperature dependence of Drude scattering time, $\tau_{Drude}$ for both MBT and MBST exhibits an initial decline up to T$\mathrm{_N}$. Here, two mechanisms largely contribute to the scattering process: (i) Drude scattering like conventional metals, which increases the scattering rate with increasing temperature and thus causes a reduction in $\tau_{Drude}$ \cite{LaForge2010,Dordevic_2013} (ii) scattering from impurity sites, which is temperature independent (nearly temperature independent Drude scattering due to the dominance of disorders, has been observed in the MBT single crystal \cite{Xu2021}). Surface electrons do not get backscattered from impurity sites because of their topological protection \cite{Sim2015}. 
But below T$\mathrm{_N}$, the gap opening of TSS results in lifting the topological protection of surface electrons from backscattering \cite{Sim2015}. Therefore below T$\mathrm{_N}$ both surface and bulk electrons follow similar scattering mechanism and show increasing scattering rate or decreasing $\tau_{Drude}$ with increasing temperature. In the region above T$\mathrm{_N}$, the topological protection is restored and hence the contribution of impurity scattering is no longer present. This can oppose the overall increase in the scattering rate. Also, in the topologically protected phase, a much lesser increase in scattering rate compared to the scattering rate of conventional metals is expected and has been reported earlier in other TIs \cite{tang2013terahertz}. These factors result in a pause in the steep declination of $\tau_{Drude}$ with increasing temperature, which decreases relatively less in the higher temperatures for MBT. In case of MBST, the bulk carrier suppression and possible enhancement of impurity or disorders (as discussed above) cause impurity scattering of surface electrons even more significant and shows a steeper declination of $\tau_{Drude}$  with increasing temperature below T$\mathrm{_N}$.  Above T$\mathrm{_N}$, $\tau_{Drude}$ for MBST remains nearly unaltered within its relatively higher error bars for the rest of the temperature range. The disorders mentioned here are mainly caused by site mixing between Mn and Bi, and are called antisite defects \cite{kopf2022,Xu2021}. Substitution of Bi via Sb has been found to increase the antisite defects substantially \cite{Riberolles2021,Lai2021}. Increased disorder because of Sb doping is consistent with the overall decrease in $\tau_{Drude}$ or an increase in the scattering rate for MBST.

\begin{figure}
\centering

     		\includegraphics[scale=0.66]{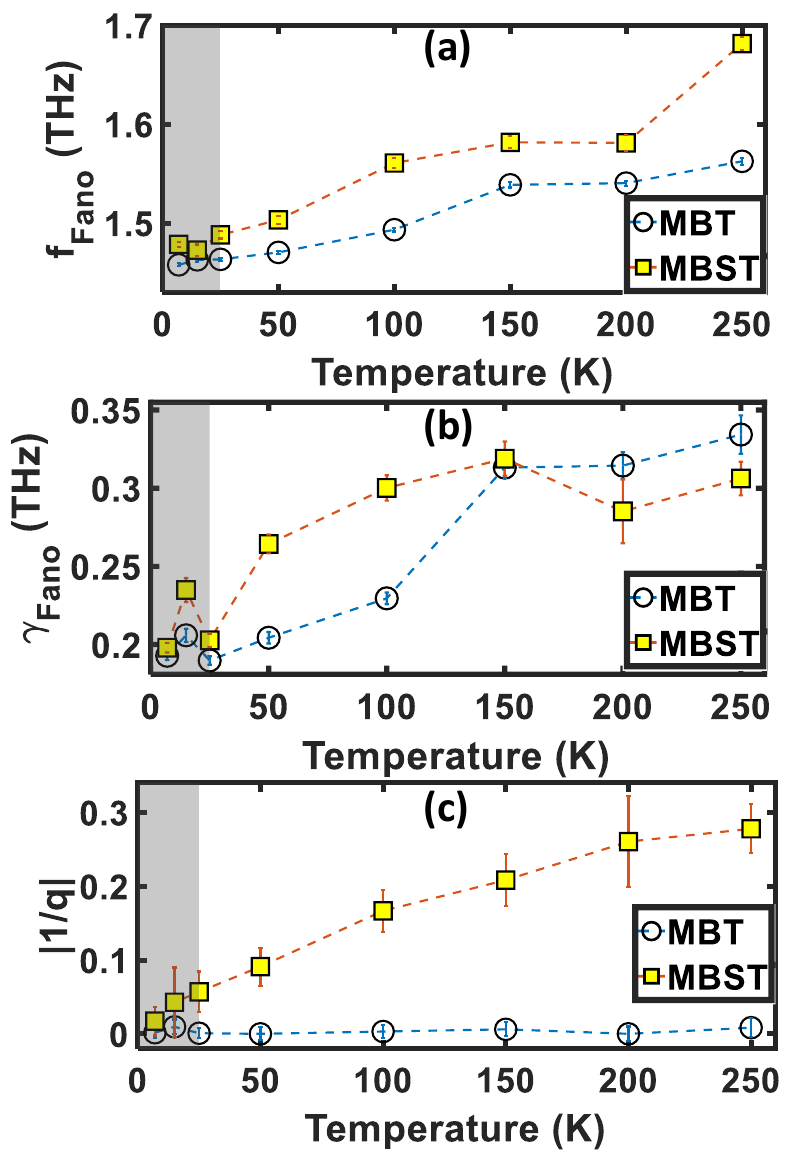}
  
   \caption{ Temperature dependence of extracted Fano parameters: (a) Frequency of the
E$_u$ phonon mode, (b) Linewidth of the
E$_u$ phonon mode and (c) Fano asymmetry parameter of the
E$_u$ phonon mode for MBT and MBST together.}
   \label{fig:5}
\end{figure}

The broader and asymmetric peak of the E$_u$ phonon mode of MBST compared to that of MBT indicates more surface electron-bulk phonon coupling in the former \cite{Choe2021}, suggesting the need of Fano oscillator for proper modelling and quantitative understanding of the electron-phonon coupling strength \cite{Sim2015}. The phonon frequency (f$_{Fano}$), linewidth ($\gamma_{Fano}$) and Fano asymmetry parameter (1/q) for MBT and MBST are shown in Fig. \ref{fig:5}(a)-\ref{fig:5}(c). For both MBT and MBST, f$_{Fano}$ shifts towards the lower frequency side with decreasing temperature. Such softening of the phonon mode upon cooling, has been seen in other TIs like Sb$_2$Te$_3$ and Bi$_2$Se$_3$ (both pristine and doped) samples \cite{li2020anomalous,wu2013sudden,bragaglia2020phonon}. The softening of the IR phonon mode, E$_u$, upon cooling is called anomalous as it is in contrast to the usual temperature dependence (hardening upon cooling) of phonons driven via the anharmonic phonon decay process and the thermal lattice expansion \cite{li2020anomalous,Kim2012,paul2021}. It is well known that some specific phonon vibrations are unstable across the phase transitions like ferroelectric \cite{ziffer2022}, magnetic \cite{Garcia2006,Gnezdilov2011}, structural \cite{choi2003}, causing their anomalous behavior with temperature. This unusual behavior has also been attributed to arise from the coupling  between the phonons and other degrees of freedom such as charge or spin \cite{paul2023,choi2003}. However in case of TIs, the coupling between surface state electrons and the bulk phonons was identified as the cause of the anomalous softening of the corresponding phonon modes \cite{melnikov2022anomalous,paul2021}. 
Exact reason for this anomalous phonon 
behavior and the influence of surface electron-bulk phonon coupling on it, will be discussed further below, followed by discussing the Fano asymmetry which is related to the electron-phonon coupling strength. The overall increase of f$_{Fano}$ in MBST compared to MBT is due to the partial substitution of Bi by the lighter atom Sb, which shifts the phonon mode to higher frequency \cite{wu2013sudden}.

Unlike the phonon frequency, the phonon-line width
for MBT shows somewhat regular behavior of increasing with temperature (Fig. \ref{fig:5}(b) ), whereas for MBST, it shows a general trend of increase, but with a seemingly sublinear behaviour. This indicates that the phonon-phonon anharmonic interaction \cite{ziffer2022,li2020anomalous,choi2003}  still plausibly  controls imaginary part of the phonon self energy (which is related to the phonon linewidth \cite{phonon_self_energy}) dominantly in case of MBT and in the case of MBST, there could be complicated contributions from other processes \cite{paul2021,kuiri2020}.

Fig. \ref{fig:5}(c) shows temperature dependence of the Fano asymmetry parameter or surface electron-bulk phonon coupling strength, $\lvert 1/q \rvert$ of MBT and MBST. 
Extremely low values of $\lvert 1/q \rvert$ ($\sim$0.005) for MBT all throughout the temperatures suggest that the Fano line-shape is not required to fit the THz conductivity. Whereas for MBST, $\lvert 1/q \rvert$ is low ($\sim$0.01) in the AFM phase. But in the paramagnetic phase (above T$\mathrm{_N}$), $\lvert 1/q \rvert$ shows considerable increment as the temperatures rises (Fig. \ref{fig:5}(c)). The reduction in Fano asymmetry from paramagnetic to AFM phase in MBST clearly suggests disappearing surface electron-bulk phonon coupling as the system undergoes magnetic ordering \cite{Choe2021,Sim2015}. A similar reduction of the Fano asymmetry in the AFM phase for a Raman active phonon mode has previously been observed in single crystalline few layer and bulk MBT \cite{Choe2021}. The absence of substantial Fano-asymmetry in the AFM phase is the consequence of the gap opening in TSS, which impacts its continuum nature as we have already discussed. This phenomena is also analogous to the reduction in electron-phonon coupling across the metal to insulator transition reported earlier in rare earth titanite, RTiO$_3$ \cite{Reedyk1997}.  For MBT, the very low value (more than ten times lower) of $\lvert 1/q \rvert$ compared to MBST even in paramagnetic phase where TSS is present, is due to the following reasons:(i) 
In case of MBT, the Fermi level is located into the conduction band (as shown in Fig. \ref{fig:1}(b)), which causes substantial filling of surface states and prevents excitation of surface electrons. Whereas, Fermi level gets shifted towards the valence band and comes closer to the Dirac node in MBST, which increases the availability of vacant states for the surface electron transitions significantly \cite{xu2017temperature}, (ii) Doping MBT with the lighter atom Sb can increase the penetration depth of the wavefunction of TSS into the bulk via modulating spin orbit coupling and bulk bandgap \cite{Sim2015,wu2013sudden}, similar to In \cite{Sim2015,wu2013sudden} and Sr \cite{melnikov2022anomalous} atom substitution in Bi$_2$Se$_3$. Therefore, MBST can host a larger spatial overlap between TSS and bulk, which causes stronger Fano interference\cite{melnikov2022anomalous,Sim2015} and results in much higher value of $\lvert 1/q \rvert$ compared to MBT.

Here, the sign of 1/q is negative which is similar to the Fano asymmetry observed in Raman phonon of bulk MBT \cite{Choe2021}. The negative sign of 1/q suggests that the bulk phonons are interacting with the surface electron absorption at frequency higher than the phonon frequency \cite{LaForge2010,Sim2015}. The increase in $\lvert 1/q \rvert$ with increasing temperature in the region above T$\mathrm{_N}$ for MBST suggests thermal excitation of carriers increases the availability of surface electrons for phonons to couple with \cite{Poojitha_2021}.

Now a closer inspection of the potential reason behind the anomalous behaviour of the E$_u$ phonon mode, which was previously attributed to its coupling with the electrons of TSS \cite{li2020anomalous,paul2021}, can be carried out. As we have discussed before, there are several factors that drive the temperature dependent frequency shift of a phonon mode. Therefore, the elucidation
of anomalous behaviour seen here need careful inspection of different contributing factors in temperature dependent phonon frequency ($\nu_{phonon}$(T)) as discussed below \cite{paul2021,saha2008}.

%\vspace{2 cm}

\begin{equation}
\begin{aligned}
    \nu_{phonon} (T)=\nu_0+\Delta \nu_{qh}(T)+\Delta \nu_{anh}(T)+\Delta \nu_{e-ph}(T)\\+\Delta\nu_{sp-ph}(T)
    \end{aligned}
\end{equation}

The first term ($\nu_0$) of the above expression is zero temperature frequency of the phonon mode. The second term ($\Delta \nu_{qh}(T)$) is the  quasiharmonic contribution due to temperature dependence of lattice constant. The third term is the anharmonic contribution of phonon decay ($\Delta \nu_{anh}(T)$) into two or more phonons. The final two terms ($\Delta\nu_{e-ph}(T)$ and $\Delta\nu_{sp-ph}(T)$) are due to modulation of phonon self energy via coupling with charge and spin, respectively \cite{paul2021,saha2008}.
The contribution of $\Delta \nu_{qh}(T)$ is generally very low \cite{bragaglia2020phonon}. Also, the thermal behaviour of lattices are generally expansive in nature (positive thermal coefficient) and therefore it decreases phonon frequency with increasing temperature unlike here \cite{Prakash2017,Kim2012}. For electron phonon coupling in TIs, phonons mostly couple with surface electron transitions \cite{melnikov2022anomalous,Sim2015} creating an electron-hole pair \cite{kuiri2020}. As discussed before, in case of MBT, the Fermi level is much above the Dirac-node of the surface state, and hence the probability of such transitions is extremely low, causing negligible electron-phonon coupling strength \cite{xu2017temperature}. This is also evident from the insignificant Fano asymmetry (1/q) seen in the phonon line-shape of MBT throughout our measured temperature range. The contribution of spin-phonon coupling can also be ignored as it appears and results in phonon softening only in the magnetically ordered phase \cite{choi2003}, unlike here. The only other channel that modulates phonon frequency or phonon self-energy is its decay into two or more phonons. The simplest decay channel is the phonon of frequency (momentum) $\nu_0 (\vec{q}_0\sim0)$ is getting decayed into two phonons, $\nu_1 (\vec{q}_1)$ and $\nu_2 (\vec{q}_2)$, following the energy ($\nu_0=\nu_1+\nu_2$) and momentum ($0=\vec{q}_1+\vec{q}_2$) conservation \cite{saha2008}. $\Delta \nu_{anh}(T)$ due to such a two phonon decay channel considering $\nu_1=\nu_2=\nu_0/2$ takes the following functional form \cite{Kim2012,saha2008,saha2009},
\begin{equation}
     \Delta \nu_{anh}(T)=A\Big(1+\frac{1}{e^{\frac{h\nu_0}{k_BT}}-1}\Big)   
\end{equation}
Here, A is the cubic anharmonicity induced self energy parameter, which is negative for normal red-shift and positive for anomalous blue-shift of the phonon frequency with increasing temperature \cite{saha2008,saha2009}. Here we observe the latter case which indicates that the restricted two-phonon density of states (integrated over all decay channels) attains its maximum at a frequency that is lower than the phonon frequency \cite{saha2008,cardona2001phonon}. In this regard, further quantitative understanding need first principle lattice dynamical calculation to be performed, which is beyond the scope of this work.

The Reason for higher anomalous shift of E$_u$ phonon frequency in MBST ($\sim$0.2 THz) in comparison to MBT ($\sim$0.1 THz) can be connected to the observation of substantially high Fano asymmetry (large values of $\lvert 1/q \rvert$). This suggests the relatively higher anomalous phonon-frequency shift in MBST can be attributed to the combined contribution of both surface electron-bulk phonon coupling ($\Delta \nu_{e-ph}$), and anomalous anharmonic phonon decay ($\Delta \nu_{anh}$). This shift is also substantially larger compared to the earlier reported anomalous shift in other TIs where only electron-phonon coupling was found to contribute \cite{melnikov2022anomalous,paul2021}.

\begin{figure}
\centering

     		\includegraphics[scale=0.73]{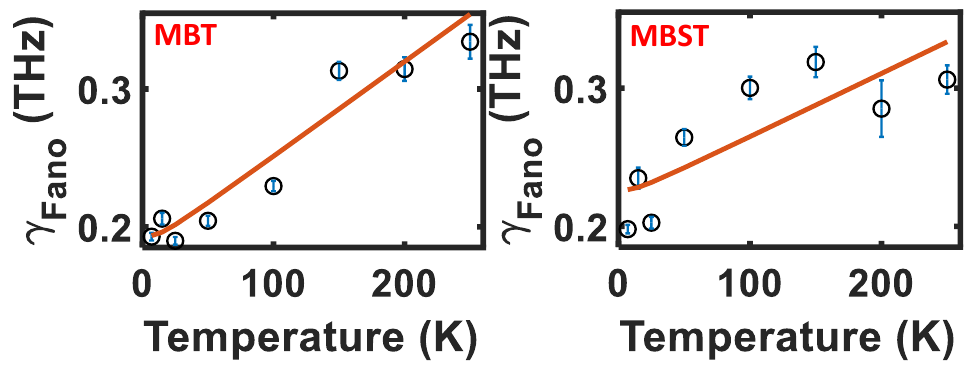}
  
   \caption{ Temperature dependent linewidth of the
E$_u$ phonon mode for (a) MBT and (b) MBST. The red continuous line is the corresponding anharmonic decay model fit using Eq. (\ref{eq:4})  }
   \label{fig:6}
\end{figure}

The temperature dependence of $\gamma_{phonon}$ for cubic anharmonicity contribution follows the following functional form,\cite{Kamaraju2010,Kamaraju2011,Prakash2017}

\begin{equation}
\gamma_{phonon} (T)=\gamma_0+C\Big(1+\frac{1}{e^{\frac{h\nu_0}{k_BT}}-1}\Big)\label{eq:4}
\end{equation}

Here, the fit parameters $\gamma_0$ and C are the disorder induced temperature independent linewidth and the measure of cubic anharmonicity, respectively. The latter is dependent on cubic phonon-phonon anharmonic interaction. The fit of $\gamma_{phonon}$ using Eq. (\ref{eq:4}) for both MBT and MBST are shown in Fig. \ref{fig:6} and the extracted parameters are listed in Supplemental Material \cite{supplemental}, Table S2. The estimated values of $\gamma_0$ for MBT and MBST are 0.182$\pm$0.023 THz and 0.219$\pm$0.039 THz, respectively. The higher value of $\gamma_0$ for MBST suggests a greater degree of disorder compared to MBT. Relatively poor fit accuracy for MBST is possibly because of electron-phonon contribution which can modulate the phonon linewidth as well \cite{paul2021} and has not been considered here, while fitting using Eq. (\ref{eq:4}).

The relatively weaker shoulder peaks of E$_u$ mode do not correspond to any theoretically predicted IR active phonon modes \cite{kobialka2022}. The presence of such unexpected shoulder peaks has previously been observed for both Raman and IR active phonon modes in different TIs \cite{li2020anomalous,Akrap2012}. The appearance of these additional peaks can be the result of disorder-induced degeneracy-lifting of the E$_u$ mode, which can cause the weak splitting of phonon modes \cite{li2020anomalous,Akrap2012}. Disorders can also cause minor misalignment within different domains and activate weak out-of-plane modes, as observed earlier in Bi$_2$Te$_2$Se \cite{Akrap2012}. Finally, attributing the origin of these extra peaks  to disorders is also consistent with the observation of more shoulder peaks in more impurity prone MBST. The fit parameters, extracted from fitting of these extra peaks are shown in Supplemental Material \cite{supplemental}, Fig. S6 and Fig. S7. These parameters have mostly shown large uncertainties unsuitable for assigning their temperature dependence to any trend, which is not surprising regarding their disorder-induced origin. 

Studies here suggest THz-TDS is an efficient spectroscopic tool for probing behaviour of surface electrons, phonons and their coupling in magnetic TIs. Furthermore, modulation of Fano asymmetry via doping and as a function of temperature has also been realized here. Thus, our investigations pave the way for fabricating a new class of quantum devices using MnBi$_2$Te$_4$, where both tunable Fano resonance and magnetic order influenced topological properties can be utilized.

\section{\label{sec:level4}conclusion}
In conclusion, we have thoroughly investigated the temperature dependence of THz conductivity of MnBi$_2$Te$_4$ and Sb doped MnBi$_2$Te$_4$ thin films, which have indicated the influence of magnetic ordering on THz response of these systems. The extracted real THz conductivity of the samples have captured a strong absorption peak of the E$_u$ phonon mode, along with the Drude-like background contributions from bulk and surface electrons. We have modelled the conductivity spectrum using the Drude-Fano-Lorentz oscillator model. The extracted plasma frequency and Drude scattering time have shown  significant changes in their temperature dependence around the magnetic ordering temperature of $\sim$25 K. The origin of these changes is attributed to the opening of an exchange gap in topological surface states in the antiferromagnetic phase. With increasing temperature (from 7K to 250K), the  E$_u$ phonon mode shows an anomalous blue-shift of $\sim$ 0.1 THz ($\sim$7\%) for MnBi$_{2}$Te$_{4}$, attributed to the positive value of cubic anharmonicity induced phonon self energy parameter. The phonon-frequency of Sb doped MnBi$_{2}$Te$_{4}$ shows relatively higher anomalous shift of $\sim$ 0.2 THz ($\sim$13\%) along with substantially high Fano asymmetry in the line-shape of the phonon mode. This indicates that the higher anomalous shift of phonon frequency in Sb doped MnBi$_{2}$Te$_{4}$ plausibly arises from the combined contribution of phonon anharmonicity and electron-phonon coupling. We believe the findings here will be instrumental in understanding magnetic order induced topological systems that show tunable Fano interference and pave the way the for fabrication of novel quantum-devices.  

\begin{acknowledgments}
The authors thank the Ministry of Education (MoE), Government of India for funding and
IISER Kolkata for the infrastructural support to carry out the research. NK thanks the SERB, Govt. of India for funding through Core Research Grant CRG/2021/004885. SM and ANM thank DST-INSPIRE and IISER Kolkata respectively for their research fellowship. S. Manna and SGN thank CSIR and UGC for their research fellowship, respectively.
\end{acknowledgments}


\clearpage


\section*{Supplementary material}

# Supplemental Material:
# Investigation of magnetic order influenced phonon and electron dynamics in $MnBi_2Te_4$ and Sb doped $MnBi_2Te_4$ through THz time-domain spectroscopy

Soumya Mukherjee,[1] Anjan Kumar NM,[1] Subhadip Manna,[1] Sambhu G Nath,[1] Radha Krishna Gopal,[2,*] Chiranjib Mitra,[1] and N. Kamaraju[1,†]

[1]*Department of Physical Sciences, Indian Institute of Science Education and Research Kolkata, Nadia, 741246, West Bengal, India*

[2]*Department of Physics and Material Sciences and Engineering, Jaypee Institute of Information Technology, Sector 62, Noida, India*

## S1. SAMPLE PREPARATION

Here the epitaxial thin films of MnBi$_2$Te$_4$ (MBT) and Sb-doped MnBi$_2$Te$_4$ (MBST) were grown on sapphire (Al$_2$O$_3$) substrate using the pulsed laser deposition (PLD) technique [1]. At first, the substrates are cleaned in an ultrasonic bath via acetone and deionized water sequentially for a duration of ∼15 minutes every time. The target pallets used for MBT and MBST deposition are consisted of 99.999% pure Mn, Bi and Te in stoichiometries of 1:2:4 and Mn, Bi, Sb and Te in stoichiometries of 1:1.7:0.3:4, respectively. The films were grown through the ablation of the target pallets using a UV KrF excimer laser (wavelength of 248 nm and the pulsed width of 25 ns) operated at a laser pulse frequency of 2 Hz.

Prior to the deposition, a base pressure of $5 \times 10^{-6}$ mbar was achieved, and during the film deposition, a partial pressure of $7.5 \times 10^{-1}$ mbar was maintained by the continuous flow of the Argon gas to achieve an appropriate plume shape. The optimized conditions, including a substrate temperature of 230° C, a target-to-substrate distance of 5 cm, and a laser fluence of 1 J cm$^{-2}$, guaranteed the growth of highly c-axis oriented crystalline thin films. After the deposition, the films were annealed at the same temperature for 20 minutes to ensure its high crystallinity and reduced surface roughness.

The crystallinity and c-axis oriented growth of the films were studied using powder X-ray diffraction (RIGAKU) and their morphology was examined using scanning electron microscopy (ZEISS SUPRA 55 VP).

## S2. SAMPLE CHARACTERIZATION

The thickness of the films which are measured using cross-sectional SEM has been shown in Fig. S1. The thicknesses are found to be ∼152 nm and ∼142 nm for MBT and MBST, respectively.

The XRD patterns of MBT and MBST thin films are shown in Fig. S2(a). The sharp peaks suggests good crystallinity for both MBT and MBST films. The strongest peak around 42° is originating from sapphire (Al$_2$O$_3$) substrate.

Energy Dispersive X-Ray (EDX) measurements have been carried out to determine the chemical composition of the samples (results are shown in Fig.S2(b)). Results of EDX

* rk.gopal@mail.jiit.ac.in
† nkamaraju@iiserkol.ac.in

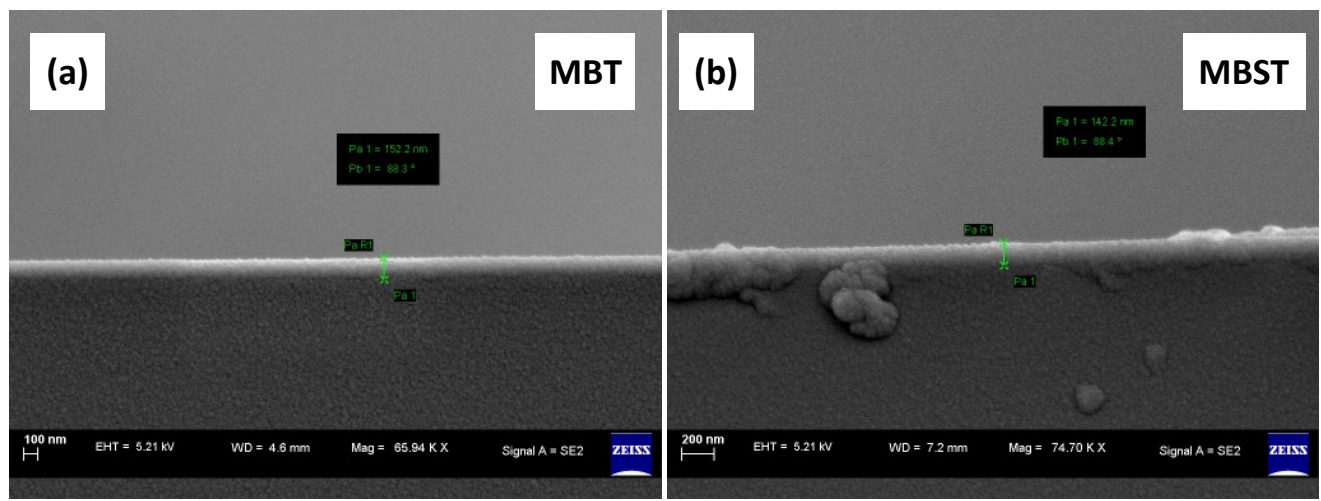

Figure S1. Cross-sectional SEM image of MBT (a) and MBST (b) thin films

spectroscopy confirms the presence of all the constituent elements for MBT. In case of MBST apart from Sb, the presence of all other constituent elements have been confirmed with adequate clarity. For both MBT and MBST the obtained Atomic percentage of Mn, Bi and Te closely matches with that of $MnBi_2Te_4$ [2].

In case of Sb, the peak is barely visible in EDX spectra due to its nominal proportion and its location in the close proximity of the Te peak [3]. Also, the proportion of Sb could not be determined from EDX spectra possibly because of the amount of Sb is below its threshold detectable amount for the EDX instrument, used here [4].

The SEM image (Fig. S2(c)) shows uniformly distributed grains of the samples with ∼100-200 nm grain size.

Raman spectra of MBT and MBST are shown in Fig. S3(a) and S3(b), respectively. Here, we can notice that apart from the Raman peaks of $MnBi_2Te_4$, there exists a few extra peaks which suggests coexistence of other polymorphs. We assign these extra peaks as the indication of $MnBi_4Te_7$ phase which is the nearest polymorph of $MnBi_2Te_4$ in $MnBi_{2n}Te_{3n+1}$ family [5, 6].

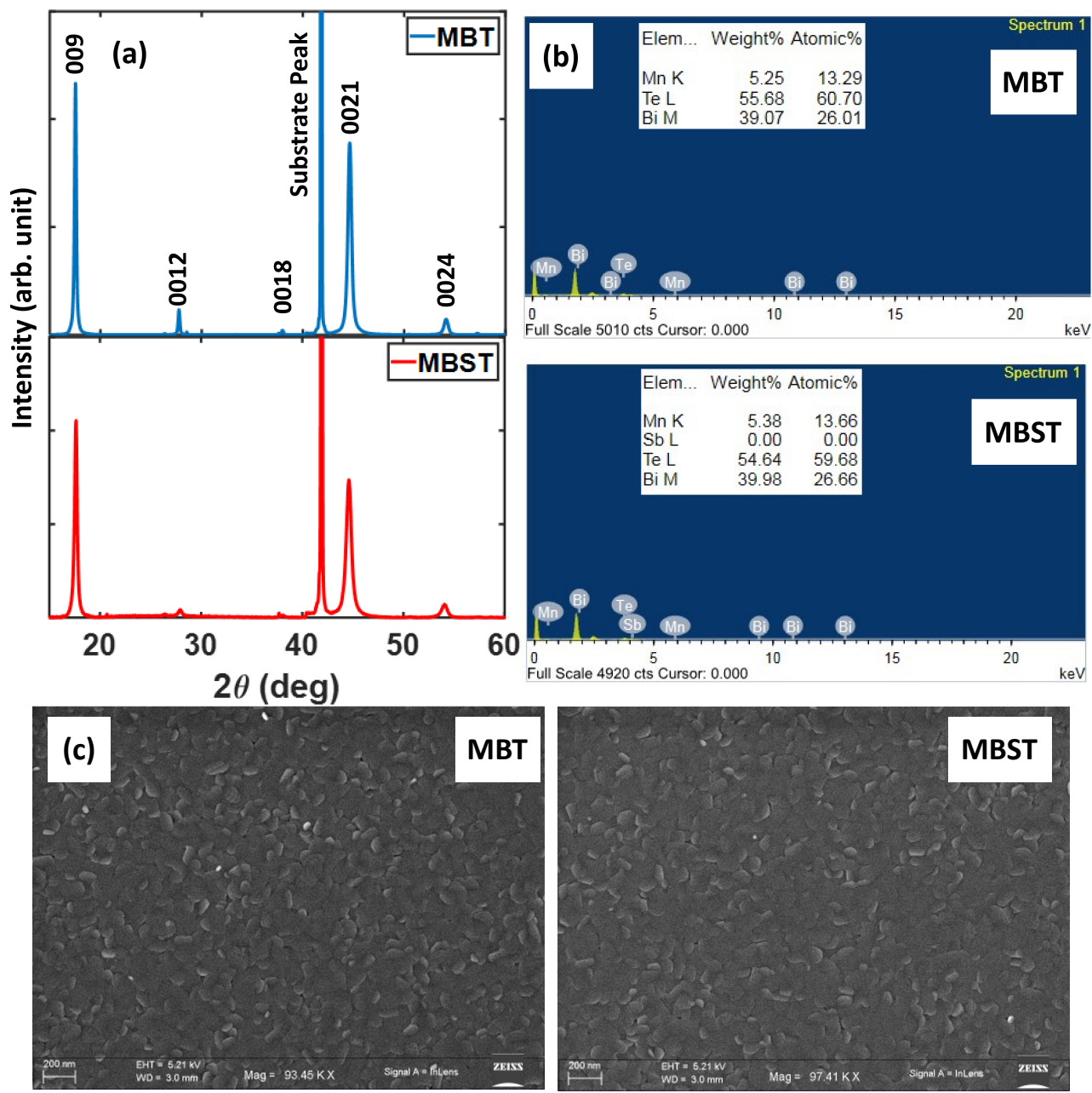

Figure S2. (a) XRD patterns (b) EDX results (c) FESEM images of MBT and MBST thin films

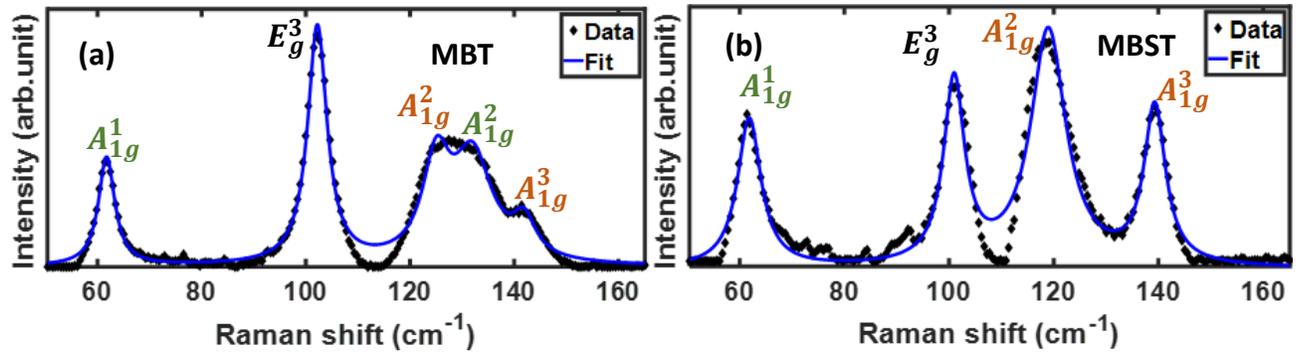

Figure S3. Raman spectra of (a) MBT and (b) MBST. The black dotted lines are measured data and continuous blue lines are fit using multiple Lorenz peaks. The peaks from $MnBi_2Te_4$ are marked by dark yellow fonts and the peaks from $MnBi_4Te_7$ are marked using green fonts and the peak which is common to both the phases is marked using black fonts.

## S3. EXPERIMENTAL DETAILS

### A. THz-TDS setup

The schematic of our home-built transmission geometry THz TDS setup, is shown in Fig. S4. The laser source used here is an amplifier pulsed-laser system (RegA 9050, Coherent Inc. (USA)) which is operated at a repetition rate of 100 kHz having pulse-width and central photon energy of ∼60 fs and ∼1.57 eV, respectively. The laser output of the amplifier then passes through a beam splitter, which divides it into two parts with intensity ratio of ∼ 70:30. The intense reflected beam is used for the THz generation and the weakly intense transmitted beam is used for detecting THz. After passing through a chopper and a retro-reflector (placed on top of a motorized delay stage), the THz-generation beam is focused with a lens onto an <110> oriented ZnTe crystal (10 mm×10 mm×1mm) which can produce THz pulses. The generated beam from ZnTe crystal then passes through a TPX sheet, which acts as an THz low pass filter and blocks all other IR components. After that the THz beam gets directed by four off-axis parabolic mirrors. The first parabolic mirror collimates the THz beam. The second one focuses the beam on to the sample, and the third one again collimates it. Finally, after passing through the Pellicle beam splitter, the fourth parabolic mirror focuses the THz beam on to another <110> oriented ZnTe crystal (10 mm×10 mm×1mm) which is used for THz detection. The detection beam is guided to the Pellicle beam splitter placed in between the third and fourth parabolic mirrors. The Pellicle beam splitter reflects the detection beam and make it propagate collinearly with the THz beam. The fourth parabolic mirror focuses the detection beam on to the detection ZnTe crystal along with the the THz beam. The detection beam after passing through the Wallotson prism, gets divided into two orthogonal components that are fed to a balanced photodiode assembly. In the absence of the THz beam,the quarter waveplate is kept in such a way that the detection beam becomes perfectly circularly polarized and the Wollaston prism provides two components of equal intensities. In the presence of the THz electric field, birefringence gets induced in the detection ZnTe crystal, which gives rise to ellipticity in the detection beam. Therefore, the orthogonal components of the detection beam emitted from the Wollaston prism becomes unequal and generate differential photocurrent proportional to the THz electric field when fed to the balanced photodiodes. A lock-in amplifier which is

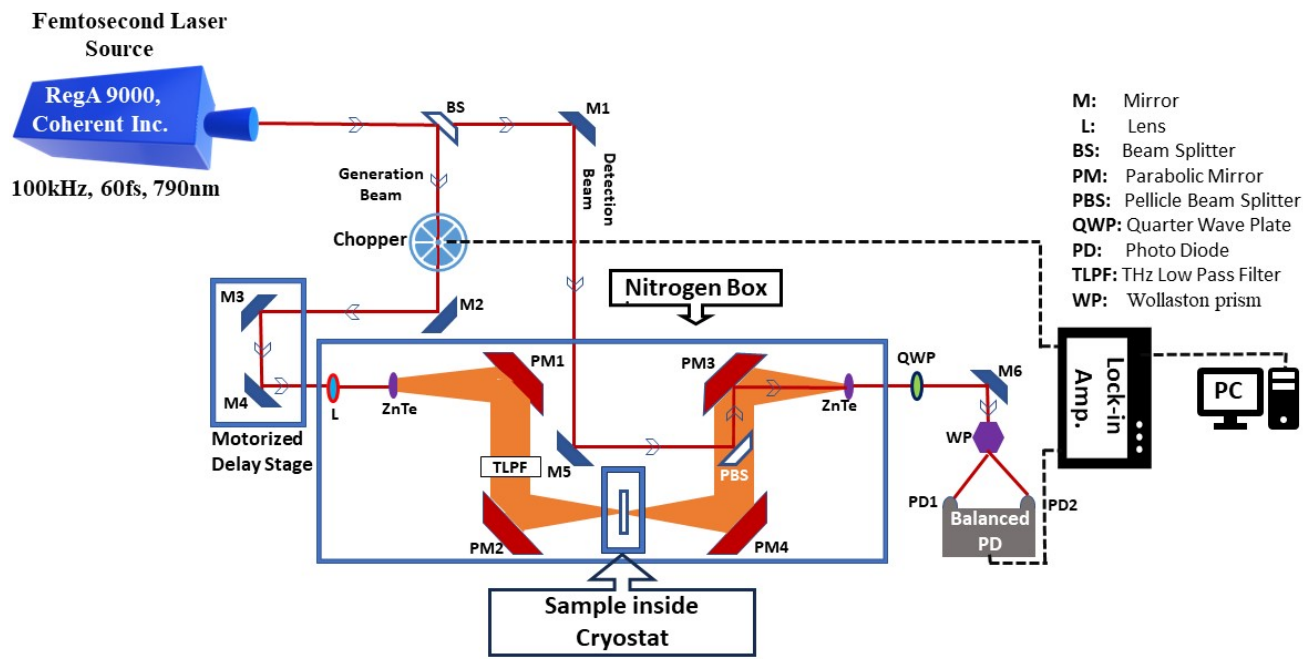

Figure S4. Schematic of the home-built THz-TDS setup used here

phase-locked to the chopper in the THz-generation path at 711 Hz measures this differential photocurrent proportional to the THz electric field. Finally, the time delay between the THz beam and the detection beam is varied through the motorized delay stage for sampling the time evolution of THz waveform in the time domain. The entire THz path is enclosed inside a sealed box and purged by 99.999% pure nitrogen gas to maintain zero humidity inside it.

## B. THz Transmission

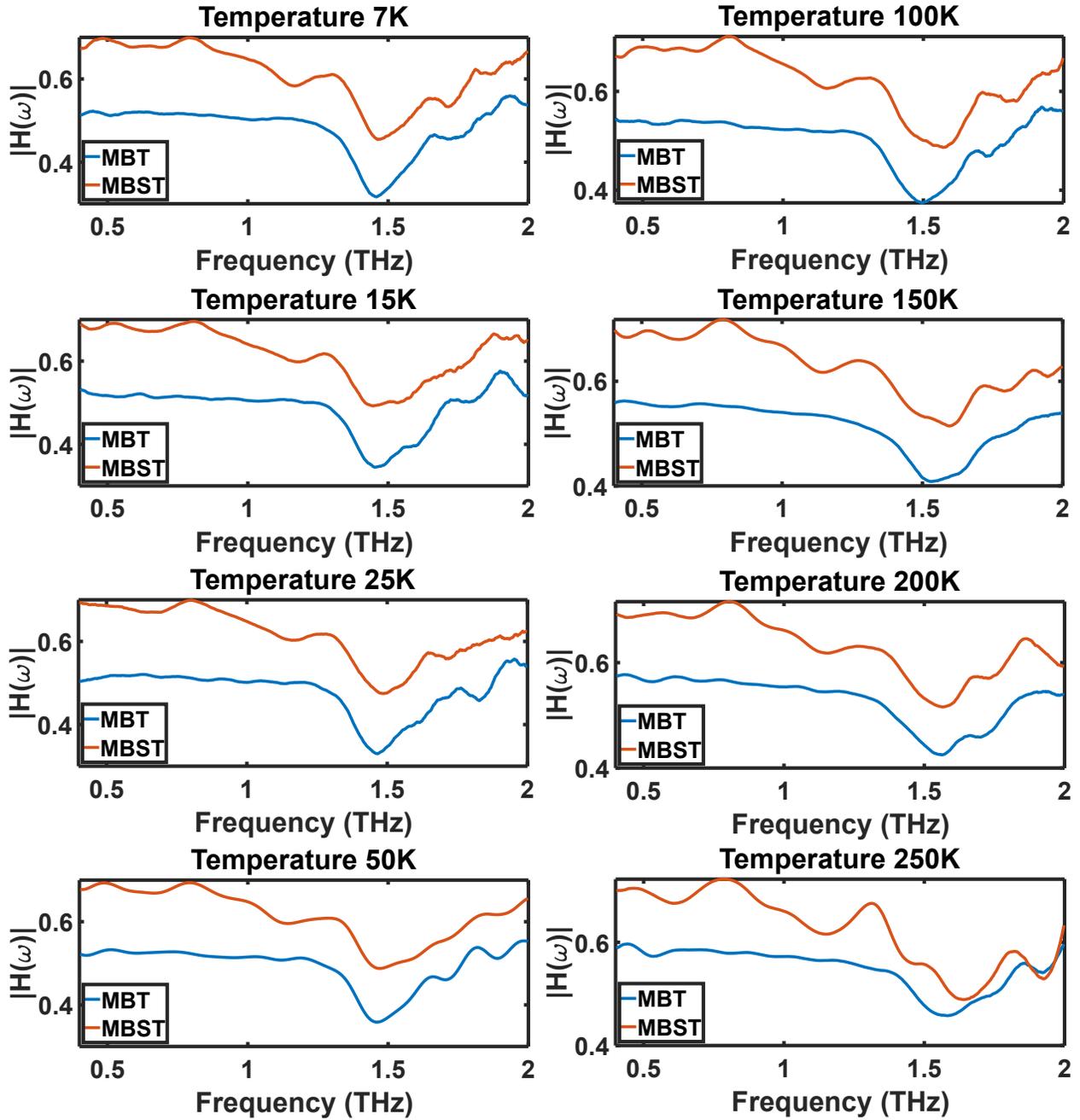

Figure S5. THz transmission ($|H(\omega)|$) at different temperatures. The temperatures are displayed on top of the each plots.

## S4. DRUDE-FANO-LORENTZ FIT OF REAL PART OF THz CONDUCTIVITY

Goodness of fit, $R^2 \left( = 1 - \frac{\sum_1 (\sigma_{real}^{exp}(i) - \sigma_{real}^{fit}(i))^2}{\sum_1 (\sigma_{real}^{exp}(i) - mean(\sigma_{real}^{exp}))^2} \right)$ values obtained from the fitting of real THz conductivity ($\sigma_{real}^{exp}$) using Eq. 1 of the main manuscript, are listed in Table S1.

Table S1. $R^2$ values of Drude-Fano-Lorentz model (Eq.1 of the main manuscript) fitting of $\sigma_{real}$ for both MBT and MBST

| MBT | | MBST | |
| --- | --- | --- | --- |
| Temperature | $R^2$ value | Temperature | $R^2$ value |
| 7 K | 0.9807 | 7 K | 0.9904 |
| 15 K | 0.9625 | 15 K | 0.9766 |
| 25 K | 0.9819 | 25 K | 0.9795 |
| 50 K | 0.9735 | 50 K | 0.9848 |
| 100 K | 0.9775 | 100 K | 0.9662 |
| 150 K | 0.9711 | 150 K | 0.9646 |
| 200 K | 0.9756 | 200 K | 0.9349 |
| 250 K | 0.9509 | 250 K | 0.9542 |

Table S2. The extracted fit parameters from cubic anharmonicity model (Eq. 4 of the main manuscript) fit of $\gamma_{Fano}$.

| Sample | $\nu_0$ (THz) | $\gamma_0$ (THz) | C (THz) |
| --- | --- | --- | --- |
| MBT | 1.425 | 0.182 | 0.012 |
| MBST | 1.425 | 0.219 | 0.008 |




\begin{thebibliography}{100}
\bibitem{phonon_self_energy}
The self-energy of phonon is a complex function, which
can be expressed as $\Sigma(\omega)$ = $\Delta(\omega)$ +i$\Gamma(\omega)$. here, the
real part, $\Delta(\omega)$ is related to the temperature dependent
shift of phonon frequency via different phonon scattering
mechanisms. Whereas the imaginary part $\Gamma(\omega)$, represents the probability of phonon decay and is proportional to the inverse of the phonon lifetime.
\end{thebibliography}
\end{document}